\documentclass[aps,preprint,superscriptaddress]{revtex4}
\usepackage{graphicx}
\usepackage{setspace}
\usepackage{amsmath}
\usepackage{dcolumn}
\usepackage{subfigure}

\newcommand{\bxi}{\mbox{\boldmath $\bf\xi$}}
\newcommand{\bfeta}{\mbox{\boldmath $\bf\eta$}}

\def\V{\textbf{V}}

\def\B{\textbf{B}}
\def\F{\textbf{F}}

\def\e{\textbf{e}}
\def\pa{\partial}
\def\R{\textbf{r}}

\begin{document}

\title{Energy of eigen-modes in magnetohydrodynamic flows of ideal fluids}

\author{I. V. Khalzov}
\affiliation{University of Saskatchewan, 116 Science Place,
Saskatoon, Saskatchewan, S7N5E2, Canada} \affiliation{Russian
Research Center "Kurchatov Institute", 1 Kurchatov Sq., Moscow,
123182, Russia.}
\author{A. I. Smolyakov}
\affiliation{University of Saskatchewan, 116 Science Place,
Saskatoon, Saskatchewan, S7N5E2, Canada} \affiliation{Russian
Research Center "Kurchatov Institute", 1 Kurchatov Sq., Moscow,
123182, Russia.}
\author{V. I. Ilgisonis}
\affiliation{Russian Research Center "Kurchatov Institute", 1
Kurchatov Sq., Moscow, 123182, Russia.}

\date{\today}

\begin{abstract}
Analytical expression for energy of eigen-modes in
magnetohydrodynamic flows of ideal fluids is obtained. It is shown
that the energy of unstable modes is zero, while the energy of
stable oscillatory modes (waves) can assume both positive and
negative values. Negative energy waves always correspond to
non-symmetric eigen-modes -- modes that have a component of
wave-vector along the equilibrium velocity. These results suggest
that all non-symmetric instabilities in ideal MHD systems with flows
are associated with coupling of positive and negative energy waves.
As an example the energy of eigen-modes is calculated for
incompressible conducting fluid rotating in axial magnetic field.
\end{abstract}

\maketitle

Energy consideration is of primary significance in stability
analysis of different magnetohydrodynamic (MHD) systems. It is well
known that the energy associated with the waves (purely oscillatory
eigen-modes) may change its sign and become negative
\cite{Fabr,Zhang}. The energy should be withdrawn from the system to
let  the negative energy wave be excited.
So, a negative energy wave is a potential source of instability
since no extra energy is needed to increase its intensity.
Instability can
arise, for example, if a negative energy wave is subject to external
dissipation; then the subsequent removal of energy from the wave
will cause it to grow. In a conservative system, the instability can
occur due to the simultaneous excitation of positive and negative
energy waves. In this case, energy is transferred from the negative
energy wave to the positive energy wave, allowing both modes to grow
and the total energy to remain constant. Waves having energies of
various signs enable researches to explain different types of
instabilities in fluid dynamics \cite{Lash}.

In the present paper we calculate the energy of the eigen-modes in
ideal one-fluid MHD and show that all instabilities of non-symmetric
eigen-modes in MHD systems with equilibrium flow are related to the
coupling of negative and positive energy waves. Following Ref.
\cite{FR}, we consider linearized dynamics of displacement vector
$\bxi$
\begin{equation}  \label{lin}
\rho\frac{\pa^2\bxi}{\pa t^2} +
2\rho(\V\cdot\nabla)\frac{\pa\bxi}{\pa t} - \F[\bxi] = 0,
\end{equation}
where $\rho$ and $\V$ are stationary values of  fluid density and
velocity, respectively. The general form of linearized force
operator $\F[\bxi]$ in ideal compressible MHD is
\begin{eqnarray}  \label{F} \F[\bxi] &=&
-\rho(\V\cdot\nabla)^2\bxi + \rho(\bxi\cdot\nabla)(\V\cdot\nabla)\V
+\nonumber\\
&+&\nabla\cdot(\rho\bxi)(\V\cdot\nabla)\V - \nabla\delta P + \\
&+&\frac{1}{4\pi}\, (\nabla\times\delta\B)\times\B +
\frac{1}{4\pi}\, (\nabla\times\B)\times\delta\B.\nonumber
\end{eqnarray}
Here, $\B$ is equilibrium magnetic field and
$$
\delta\B = \nabla\times(\bxi\times\B)
$$
is its perturbation. The perturbation of fluid pressure $\delta P$
can be specified  by thermodynamic properties of the system. For
example, if the process is adiabatic with adiabatic index $\gamma$
then
$$
\delta P = -\bxi\cdot\nabla P - \gamma P\nabla\cdot\bxi.
$$
In the case of incompressible MHD, such equation appears to be
excessive, instead one has to impose the incompressibility condition
$\nabla\cdot\bxi=0$.

A number of formal properties of Eq. (\ref{lin}) can be established.
Force operator $\F[\bxi]$ is Hermitian (self-adjoint) in the
following sense,
\begin{equation}
\label{prop1} \int\bfeta\cdot\F[\bxi]\,d^3\R =
\int\bxi\cdot\F[\bfeta]\, d^3\R,
\end{equation}
while the second term in Eq. (\ref{lin}) is antisymmetric:
\begin{equation}
\label{prop2} \int\rho\bfeta\cdot(\V\cdot\nabla)\bxi \, d^3\R =
-\int\rho\bxi\cdot(\V\cdot\nabla)\bfeta \, d^3\R.
\end{equation}
Integration in Eqs. (\ref{prop1}) and (\ref{prop2}) is performed
over the fluid volume under the assumption that displacements on the
boundary vanish.

In our subsequent discussion, the displacement vector $\bxi$ is
supposed to be complex. In order to obtain the correct expression
for energy of perturbations  in this case, we multiply Eq.
(\ref{lin}) by complex conjugate $\pa\bxi^*/\pa t$  and integrate
over the space:
\begin{equation}  \label{com1}
\int\bigg(\rho\frac{\pa\bxi^*}{\pa t}\cdot\frac{\pa^2\bxi}{\pa t^2}
+ 2\rho\frac{\pa\bxi^*}{\pa t}\cdot(\V\cdot\nabla)\frac{\pa\bxi}{\pa
t} - \frac{\pa\bxi^*}{\pa t}\cdot\F[\bxi]\bigg)\,d^3\R = 0.\nonumber
\end{equation}
The complex conjugate of this equality is
\begin{equation}  \label{com2}
\int\bigg(\rho\frac{\pa\bxi}{\pa t}\cdot\frac{\pa^2\bxi^*}{\pa t^2}
+ 2\rho\frac{\pa\bxi}{\pa t}\cdot(\V\cdot\nabla)\frac{\pa\bxi^*}{\pa
t} - \frac{\pa\bxi}{\pa t}\cdot\F[\bxi^*]\bigg)\,d^3\R = 0.\nonumber
\end{equation}
Summing up these two equations
and using the properties (\ref{prop1}), (\ref{prop2}) we arrive at
the energy conservation law in the form $\pa E/\pa t=0$, where
\begin{equation}  \label{E}
E=\frac{1}{2}\int\bigg(\rho\bigg|\frac{\pa\bxi}{\pa t}\bigg|^2 -
\bxi^*\cdot\F[\bxi]\bigg)\,d^3\R.
\end{equation}
As usual in mechanics, the total energy of the perturbations
consists of kinetic part (first term) and of potential part (second
term).

Since the equilibrium quantities have no time dependence, we look
for a normal-mode solutions to Eq. (\ref{lin}) in the form
\begin{equation}\label{mode}
\bxi(\R,t)=\hat{\bxi}(\R)e^{-i\omega t}.
\end{equation}
Then, the equation of motion (\ref{lin}) leads to eigen-value
problem
\begin{equation}  \label{eigen}
\omega^2\rho\hat{\bxi} + 2i\omega\rho(\V\cdot\nabla)\hat{\bxi} +
\F[\hat{\bxi}]=0.
\end{equation}
Multiplying this equation by complex conjugate $\hat{\bxi}^*$ and
integrating over the fluid volume, we arrive at quadratic equation
for eigen-frequency $\omega$,
\begin{equation}  \label{quad}
A\,\omega^2- 2B\,\omega - C =0,
\end{equation}
with coefficients
\begin{eqnarray}
  A &=& \int\rho|\hat{\bxi}|^2\,d^3\R>0,\nonumber\\
  B &=& -\,i\int\rho\hat{\bxi}^*\cdot(\V\cdot\nabla)\hat{\bxi}\,d^3\R,\nonumber\\
  C &=& -\int\hat{\bxi}^*\F[\hat{\bxi}]\,d^3\R.\nonumber
\end{eqnarray}
Solving Eq. (\ref{quad}) we find
\begin{equation}  \label{om}
\omega_{1,2} = \frac{B\pm\sqrt{B^2+AC}}{A}.
\end{equation}
This expression allows to determine eigen-frequency corresponding to
known eigen-mode $\hat{\bxi}$. Since all coefficients in Eq.
(\ref{quad}) are real [for coefficients $C$ and $B$ it follows
immediately from properties (\ref{prop1}) and (\ref{prop2}),
respectively], the instability in the system is possible if and only
if $B^2+AC<0$ for some eigen-mode.

Now we are able to determine the energy of the eigen-mode with
eigen-frequency (\ref{om}). Substituting (\ref{mode}) into
expression (\ref{E}) we obtain:
\begin{equation}  \label{modeEn}
E = \frac{1}{2}\,(A\,|\,\omega|^2 + C).
\end{equation}
In the case of unstable mode, $B^2+AC<0$, so
$$
|\,\omega_{1,2}|^2=-\frac{C}{A},
$$
and the energy is
\begin{equation}  \label{EnUnst}
E_{1,2} = 0.
\end{equation}
For stable mode, $B^2+AC\geq0$ and the energy is
\begin{eqnarray}  \label{EnSt}
E_{1,2} &=& \frac{\sqrt{B^2+AC}}{A}\,\bigg(\sqrt{B^2+AC} \pm B
\bigg)=\\
&=& \pm\,\omega\,\sqrt{B^2+AC}.
\end{eqnarray}
Therefore, energy of stable eigen-mode changes the sign if its
frequency changes the sign.

Depending on the system parameters different options are realized in
the case of stable eigen-modes (Table \ref{t1}). As one can see,
there is an interval of parameters at which the waves with positive
and negative energy coexist (option 2). One boundary of this
interval corresponds to the stability threshold (option 1), the
other -- to change of sign of eigen-frequency $\omega_2$ (option 3).
This result suggests that the instability in the ideal MHD system
with flow can be associated with coupling of positive and negative
energy waves.

We note here that all negative energy waves are non-symmetric modes,
i.e., they have spatial dependence along the equilibrium flow, so,
the coefficient $B\ne0$. For symmetric modes or in the absence of
flow we have $B=0$ and the energy is
$$
E=\frac{1}{2}\,(|C|+C).
$$
Therefore, energy of symmetric modes is never negative, and their
stability can be investigated by use of energy principle \cite{FR}.
In a case of non-axisymmetric modes, the energy principle fails and
special arrangements should be made to modify it (see, e.g.,
\cite{myJETP}).

\begin{table}[tb]
\caption{Eigen-frequencies $\omega_{\,1,2}$ and corresponding
energies $E_{\,1,2}$ of stable eigen-modes for different values of
coefficient $C$ ($B\geq0$ is assumed for simplicity). \label{t1}}
\begin{ruledtabular}
\begin{tabular}{cccccc}
& $C$ & $\omega_1$ & $E_1$ & $\omega_2$ & $E_2$ \\
\hline\\
1. & $-B^2/A$ & + & 0 & + & $0$\\
2. & $(-B^2/A;~0)$ & + & + & + & $-$\\
3. & $0$ & + & + & 0 & $0$\\
4. & $(0;~\infty)$ & + & + & $-$ & +\\
\end{tabular}
\end{ruledtabular}
\end{table}

In order to verify the above analytical results, we calculate the
energy of eigen-modes of incompressible fluid rotating in homogenous
transverse magnetic field $\B=B\e_z$. The equilibrium velocity
profile used in our calculations corresponds to the electrically
driven flow in circular channel and has a form
\begin{equation}\label{Omega}
\V=r\Omega(r)\e_\varphi,~~~\Omega(r)=\frac{\Omega_1r_1^2}{r^2}
\end{equation}
in cylindrical system of coordinates $\{r,\varphi,z\}$. Here, $r_1$
and $r_2$ are inner and outer radii of the channel, respectively,
and $\Omega_1$ is the angular velocity at $r_1$. This type of flow
profile is used in new experimental device \cite{RRC} for laboratory
testing of the so-called magnetorotational instability (MRI), which
plays an important role in many astrophysical applications (see
reviews \cite{BH_rev,Balbus_rev}).

A detailed eigen-mode analysis of such flow has been performed in
Ref. \cite{myPF}. Assuming
$$
\bxi(\R,t)=\bxi(r)e^{-i\omega t + im\varphi + ik_zz}
$$
one obtains eigen-value problem
\begin{eqnarray}\label{EVP}
&&(\bar{\omega}^2-\omega_A^2)\bxi
+2i\Omega\bar{\omega}(\xi_r\e_\varphi-\xi_\varphi\e_r) -
\frac{\pa\Omega^2}{\pa r}\,r\xi_r\e_r = \nabla\delta\Pi,\nonumber\\
&&\frac{1}{r}\frac{\pa(r\xi_r)}{\pa r} + \frac{im}{r}\,\xi_\varphi +
ik_z\xi_z=0
\end{eqnarray}
with boundary conditions
\begin{equation}\label{bc}
\xi_r(r_1)=\xi_r(r_2)=0,
\end{equation}
where
$$
\omega_A=\frac{k_zB}{\sqrt{4\pi\rho}}
$$
is Alfven frequency,
$$
\bar{\omega}=\omega-m\Omega
$$
is "shifted"\ eigen-frequency and $\delta\Pi$ is perturbation of the
total normalized pressure,
$$
\Pi=\frac{P}{\rho} + \frac{\B^2}{4\pi\rho}.
$$
A general expression for energy of perturbations (\ref{E}) for this
system reads
\begin{eqnarray}  \label{EnED}
E=\frac{\rho}{2}\int\bigg(\bigg|\frac{\pa\bxi}{\pa t}\bigg|^2 &+&
(\omega_A^2-m^2\Omega^2)|\bxi|^2 + r\frac{\pa\Omega^2}{\pa
r}\,|\,\xi_r|^2+\nonumber\\
&+& 2im\Omega^2(\xi_r\xi^*_{\varphi} - \xi^*_r\xi_{\varphi})
\bigg)\,d^3\R.
\end{eqnarray}
Substituting $\bxi$ from the eigen-value problem (\ref{EVP}),
(\ref{bc}) we find the energy of stable mode with frequency
$\omega$:
\begin{eqnarray}  \label{EnED2}
E&=&\pi\rho h\omega\int\limits_{r_1}^{r_2}r\bigg\{\bar{\omega}
\bigg[ \xi_r^2 + \frac{1}{m^2+k_z^2r^2}\bigg(\frac{\pa(r\xi_r)}{\pa
r}\bigg)^2 +\\
&+&\frac{4k_z^2r^2\Omega^2\omega_A^2\xi_r^2}{(\omega_A^2-\bar{\omega}^2)^2(m^2+k_z^2r^2)}
\bigg]+\frac{2m\Omega\xi_r}{m^2+k_z^2r^2}\frac{\pa(r\xi_r)}{\pa r}
\bigg\}\,dr,\nonumber
\end{eqnarray}
where $h$ is the height of the channel. For axisymmetric eigen-modes
with $m=0$ this expression is reduced to
\begin{equation}  \label{EnED3}
E=\pi\rho h\omega^2\int\limits_{r_1}^{r_2}\bigg[\xi_r^2 +
\frac{1}{k_z^2r^2}\bigg(\frac{\pa(r\xi_r)}{\pa r}\bigg)^2 +
\frac{4\Omega^2\omega_A^2\xi_r^2}{(\omega_A^2-\omega^2)^2}
\bigg]r\,dr.\nonumber
\end{equation}
Therefore, the energy of axisymmetric eigen-modes is always positive
if $\omega\ne0$. Formally, this case is described by (\ref{EnSt})
with coefficient $B=0$.

\begin{figure}[tb]
\includegraphics[scale=0.5]{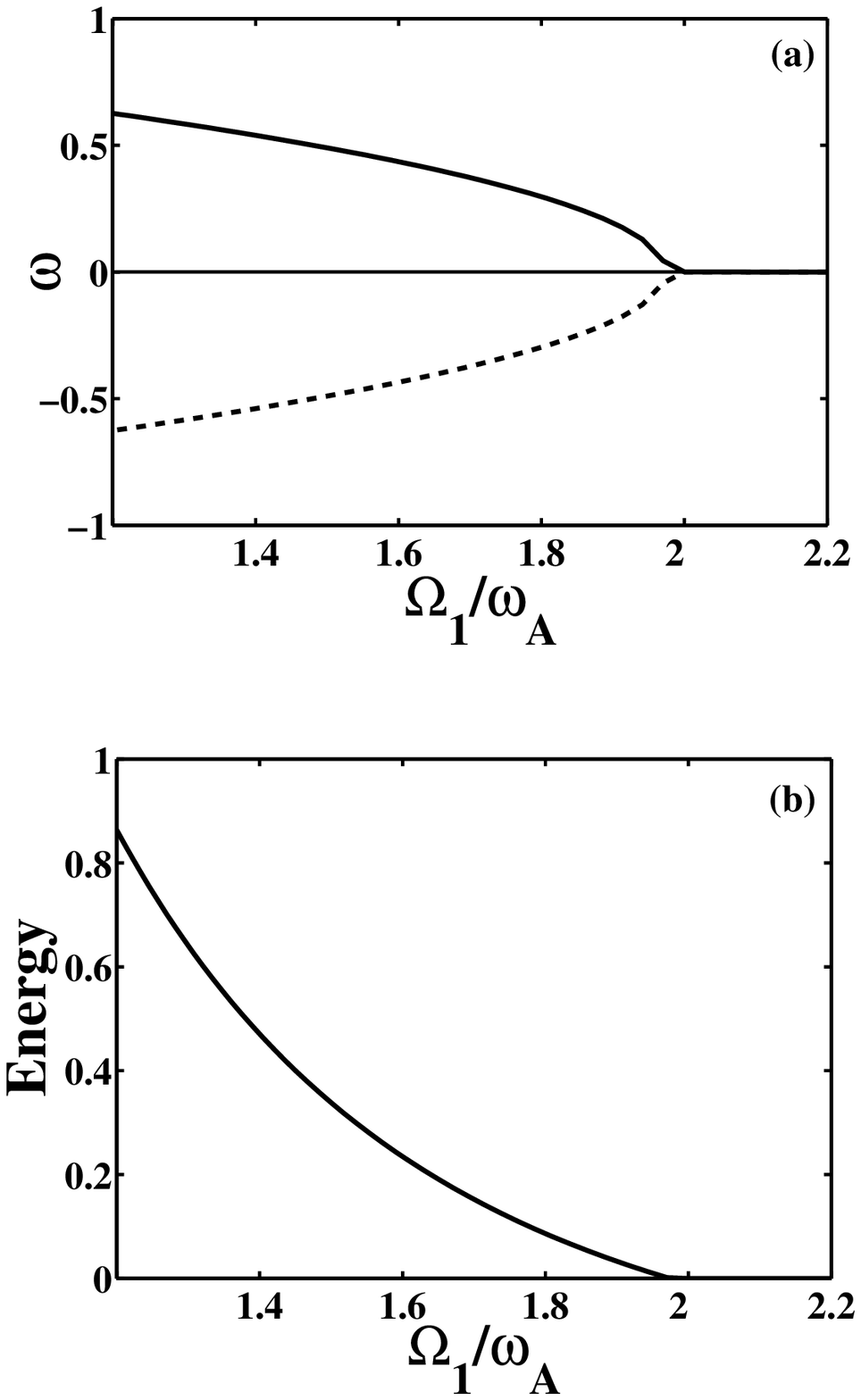}
\caption{Calculated dependence of eigen-frequency (a) and energy (b)
on ratio $\Omega_1/\omega_A$ for two most unstable eigen-modes with
azimuthal number $m=0$. Energy is given in arbitrary units.}
\label{m0}
\end{figure}
\begin{figure}[tb]
\includegraphics[scale=0.5]{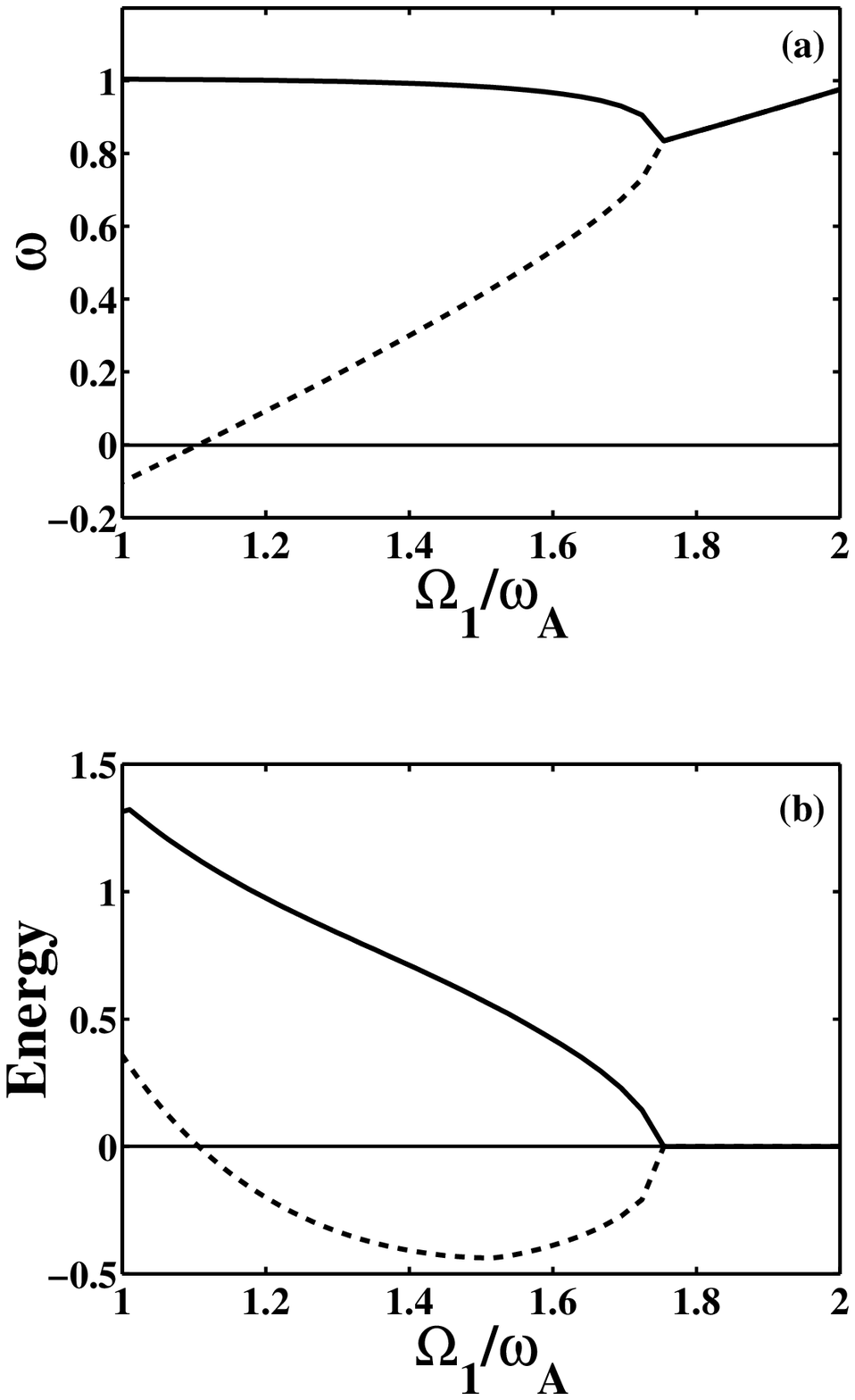}
\caption{Calculated dependence of eigen-frequency (a) and energy (b)
on ratio $\Omega_1/\omega_A$ for two most unstable eigen-modes with
azimuthal number $m=1$. Energy is given in arbitrary units.}
\label{m1}
\end{figure}

In Figs. \ref{m0}, \ref{m1} the calculated dependencies of frequency
and energy  for two potentially unstable eigen-modes on the
parameter $\Omega_1/\omega_A$ are shown. In the axisymmetric case
($m=0$), both branches of energy are positive and coincident (Fig.
\ref{m0}b). The merging point in Fig. \ref{m0}a corresponds to
$\Omega_1/\omega_A\approx2.0$ which is the threshold of
magnetorotational instability for $m=0$.
The nature of axisymmetric MRI is not related to the subject of
negative energy waves and can be explained by the mechanism similar
to one of Raleigh-Taylor instability \cite{Vel}.

For $m=1$ the behavior of both energy curves in Fig. \ref{m1}b is
completely described by Eq. (\ref{EnSt}). The MRI threshold in this
case is $\Omega_1/\omega_A\approx1.7$.
When $1.1\lesssim\Omega_1/\omega_A\lesssim1.7$ the positive and
negative energy waves can coexist in the system. At
$\Omega_1/\omega_A\approx1.1$ the frequency $\omega_2$ changes the
sign (Fig. \ref{m1}a, dashed line), so both energy branches become
positive.

\begin{figure}[tb]
\includegraphics[scale=0.3]{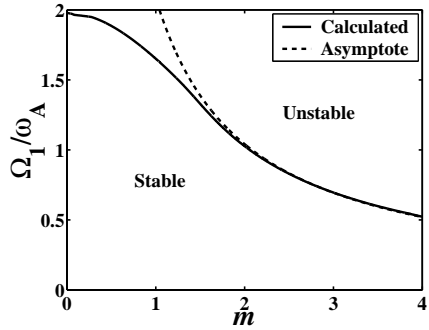}
\caption{Calculated dependence of magnetorotational instability
threshold on azimuthal mode number $m$ (solid line) and its
asymptote for large $m$ (dashed line).} \label{Om_m}
\end{figure}

It should be noted that merging points in Figs. \ref{m0} and
\ref{m1} determine the magnetorotational instability threshold. In
the flow given by (\ref{Omega}) this threshold decreases with
azimuthal number $m$, as discussed in Ref. \cite{myPF}. For large
$m$ it approaches the asymptote
\begin{equation}\label{MS}
\frac{\Omega_1}{\omega_A}=\frac{2}{m(1-r_1^2/r_2^2)}.
\end{equation}
The calculated dependence of MRI threshold on small $m$ is presented
in Fig. \ref{Om_m}.

In conclusion, we have shown that all non-symmetric MHD
instabilities in ideal fluids with flows can be explained as a
coupling of originally stable positive and negative energy waves.
These results are supported by calculations of frequencies and
energies of eigen-modes in the flow that can be unstable with
respect to magnetorotational instability.

This work is supported in part by NSERC Canada.

\newpage

\bibliography{Negative}

\begin{thebibliography}{10}
\expandafter\ifx\csname natexlab\endcsname\relax\def\natexlab#1{#1}\fi
\expandafter\ifx\csname bibnamefont\endcsname\relax
  \def\bibnamefont#1{#1}\fi
\expandafter\ifx\csname bibfnamefont\endcsname\relax
  \def\bibfnamefont#1{#1}\fi
\expandafter\ifx\csname citenamefont\endcsname\relax
  \def\citenamefont#1{#1}\fi
\expandafter\ifx\csname url\endcsname\relax
  \def\url#1{\texttt{#1}}\fi
\expandafter\ifx\csname urlprefix\endcsname\relax\def\urlprefix{URL }\fi
\providecommand{\bibinfo}[2]{#2}
\providecommand{\eprint}[2][]{\url{#2}}

\bibitem[{\citenamefont{Fabrikant and Stepanyants}(1998)}]{Fabr}
\bibinfo{author}{\bibfnamefont{A.~L.} \bibnamefont{Fabrikant}}
  \bibnamefont{and} \bibinfo{author}{\bibfnamefont{Y.~A.}
  \bibnamefont{Stepanyants}}, \emph{\bibinfo{title}{Propagation of waves in
  shear flows}} (\bibinfo{publisher}{World Scientific}, \bibinfo{year}{1998}).

\bibitem[{\citenamefont{Zhang and Lovelace}(2005)}]{Zhang}
\bibinfo{author}{\bibfnamefont{L.}~\bibnamefont{Zhang}} \bibnamefont{and}
  \bibinfo{author}{\bibfnamefont{R.~V.~E.} \bibnamefont{Lovelace}},
  \bibinfo{journal}{Astrophys. and Space Science}
  \textbf{\bibinfo{volume}{300}}, \bibinfo{pages}{395} (\bibinfo{year}{2005}).

\bibitem[{\citenamefont{Lashmore-Davies}(2005)}]{Lash}
\bibinfo{author}{\bibfnamefont{C.~N.} \bibnamefont{Lashmore-Davies}},
  \bibinfo{journal}{J. Plasma Phys.} \textbf{\bibinfo{volume}{71}},
  \bibinfo{pages}{101} (\bibinfo{year}{2005}).

\bibitem[{\citenamefont{Frieman and Rotenberg}(1960)}]{FR}
\bibinfo{author}{\bibfnamefont{E.}~\bibnamefont{Frieman}} \bibnamefont{and}
  \bibinfo{author}{\bibfnamefont{M.}~\bibnamefont{Rotenberg}},
  \bibinfo{journal}{Rev. Mod. Phys.} \textbf{\bibinfo{volume}{32}},
  \bibinfo{pages}{898} (\bibinfo{year}{1960}).

\bibitem[{\citenamefont{Ilgisonis and Khalzov}(2005)}]{myJETP}
\bibinfo{author}{\bibfnamefont{V.~I.} \bibnamefont{Ilgisonis}}
  \bibnamefont{and} \bibinfo{author}{\bibfnamefont{I.~V.}
  \bibnamefont{Khalzov}}, \bibinfo{journal}{JETP Letters}
  \textbf{\bibinfo{volume}{82}}, \bibinfo{pages}{570} (\bibinfo{year}{2005}).

\bibitem[{\citenamefont{Velikhov et~al.}(2006)\citenamefont{Velikhov, Ivanov,
  Zakharov, Zakharov, Livadny, and Serebrennikov}}]{RRC}
\bibinfo{author}{\bibfnamefont{E.~P.} \bibnamefont{Velikhov}},
  \bibinfo{author}{\bibfnamefont{A.~A.} \bibnamefont{Ivanov}},
  \bibinfo{author}{\bibfnamefont{S.~V.} \bibnamefont{Zakharov}},
  \bibinfo{author}{\bibfnamefont{V.~S.} \bibnamefont{Zakharov}},
  \bibinfo{author}{\bibfnamefont{A.~O.} \bibnamefont{Livadny}},
  \bibnamefont{and} \bibinfo{author}{\bibfnamefont{K.~S.}
  \bibnamefont{Serebrennikov}}, \bibinfo{journal}{Physics Letters A}
  \textbf{\bibinfo{volume}{358}}, \bibinfo{pages}{216} (\bibinfo{year}{2006}).

\bibitem[{\citenamefont{Balbus and Hawley}(1998)}]{BH_rev}
\bibinfo{author}{\bibfnamefont{S.~A.} \bibnamefont{Balbus}} \bibnamefont{and}
  \bibinfo{author}{\bibfnamefont{J.~F.} \bibnamefont{Hawley}},
  \bibinfo{journal}{Rev. Mod. Phys.} \textbf{\bibinfo{volume}{70}},
  \bibinfo{pages}{1} (\bibinfo{year}{1998}).

\bibitem[{\citenamefont{Balbus}(2003)}]{Balbus_rev}
\bibinfo{author}{\bibfnamefont{S.~A.} \bibnamefont{Balbus}},
  \bibinfo{journal}{Annu. Rev. Astron. Astrophys.}
  \textbf{\bibinfo{volume}{41}}, \bibinfo{pages}{555} (\bibinfo{year}{2003}).

\bibitem[{\citenamefont{Khalzov et~al.}(2006)\citenamefont{Khalzov, Ilgisonis,
  Smolyakov, and Velikhov}}]{myPF}
\bibinfo{author}{\bibfnamefont{I.~V.} \bibnamefont{Khalzov}},
  \bibinfo{author}{\bibfnamefont{V.~I.} \bibnamefont{Ilgisonis}},
  \bibinfo{author}{\bibfnamefont{A.~I.} \bibnamefont{Smolyakov}},
  \bibnamefont{and} \bibinfo{author}{\bibfnamefont{E.~P.}
  \bibnamefont{Velikhov}}, \bibinfo{journal}{Phys. Fluids}
  \textbf{\bibinfo{volume}{18}}, \bibinfo{eid}{124107} (\bibinfo{year}{2006}).

\bibitem[{\citenamefont{Velikhov}(1959)}]{Vel}
\bibinfo{author}{\bibfnamefont{E.~P.} \bibnamefont{Velikhov}},
  \bibinfo{journal}{Sov. Phys. JETP} \textbf{\bibinfo{volume}{9}},
  \bibinfo{pages}{995} (\bibinfo{year}{1959}).

\end{thebibliography}

\end{document}